\documentclass[10pt,conference]{IEEEtran}
\IEEEoverridecommandlockouts

\usepackage{cite}
\usepackage{graphicx}
\usepackage{amsmath,amssymb,amsfonts}
\usepackage{booktabs}
\usepackage{tabularx}
\usepackage{multirow}
\usepackage{url}
\usepackage[hidelinks]{hyperref}
\usepackage{balance}
\usepackage[most]{tcolorbox}

\newtcolorbox{finding}{
  enhanced,
  colback=black!4, colframe=black!60,
  boxrule=0.3pt, leftrule=2.2pt,
  left=5pt, right=5pt, top=3pt, bottom=3pt,
  arc=0.5pt, boxsep=1pt,
  before skip=5pt, after skip=5pt
}

\title{After the Party: Growth, Governance, and Security Scanning in the OpenClaw Agent Skill Ecosystem}

\author{
  \IEEEauthorblockN{Yunpeng Xiong}
  \IEEEauthorblockA{Monash University, Australia \\
   nemo.xiong@monash.edu}
  \and
    \IEEEauthorblockN{
      Ting Zhang\textsuperscript{\ensuremath{\heartsuit}}
      \thanks{\ensuremath{\heartsuit} Ting Zhang is the corresponding author.}
    }
    \IEEEauthorblockA{Monash University, Australia \\
    ting.zhang@monash.edu}
}

\begin{document}
\bstctlcite{BSTcontrol}
\maketitle

\begin{abstract}
AI agents increasingly act through agent skills, i.e., natural-language instructions, that direct a host agent toward shell, network, credential, file, and process actions, and public registries distribute them at scale. 
In the first half of 2026, the OpenClaw AI agent went viral, and its public skill registry boomed: the observable stock nearly doubled in 91 days, and a majority of the listings visible in June were created in just two months.
By the end of our study window, the wave had crested, and monthly listing creation and core-repository activity were falling from their spring peaks.
This paper measures what the boom left behind, drawing on the OpenClaw Git history, its GitHub issues and pull requests, and three ClawHub registry snapshots.
Attention is concentrated: the top 10\% of skills received 46.93\% of all downloads. 
No simple skill features (like size or download counts) remained a stable predictor of continued listing once creation cohort and skill age were controlled.
Human scrutiny did not stay: 77.86\% have zero stars and zero comments, while 85.06\% of the readable skills carry privilege evidence.
And automated cleanup is not ready: the three security scanners disagreed on 23,702 of the 61,990 skills they all cover.
After human adjudication, weighted scanner sensitivity against the reference standard ranged from 21.67\% to 61.06\%.
Governing fast-growing agent-skill registries cannot rely on simple metadata or single scanner scores;
it requires robust, transparent measurement and independent validation.
\end{abstract}

\begin{IEEEkeywords}
agent skills, software ecosystems, software repositories, registry governance, security measurement, empirical software engineering
\end{IEEEkeywords}

\section{Introduction}
Agent skills are a new kind of software artifact: instruction manuals (typically a \texttt{SKILL.md} file) that tell an AI agent how to use tools and perform actions \cite{agentskills}. 
Unlike a conventional software library, a skill carries little or no executable code, yet within a trusted host it produces real side effects.
In early 2026, OpenClaw~\cite{openclaw2026}, an open-source AI agent, went viral, and ClawHub, the public registry where people share skills for it, boomed with it \cite{openclawRepo2026,clawhubRepo2026}: the observable stock nearly doubled in 91 days, from 33,399 to 65,175 listings, and 63.25\% of the listings visible in June were created in March and April alone.
By the end of our study window, the wave has crested: listing creation peaked in March and fell to a fraction of that peak by May, the last fully observed month, while core-repository commits fell by more than half into June.

What the boom left behind should still be governed.
In practice, a registry has three families of signals to govern with: the metadata it records (e.g., downloads, stars, versions), the community feedback it displays, and the verdicts of the automated scanners it runs.
Yet every one of these signals was minted while the population doubled within a quarter, and none has been validated: prior work measures agent skills at scale~\cite{liu2026agent,ling2026agent}, analyzes OpenClaw-specific threats~\cite{deng2026taming, shan2026openclaw, suwansathit2026openclaw}, and benchmarks skill organization~\cite{li2026agentskillos, liu2026skillusage, li2026skillsbench}, but no study jointly examines an ecosystem's scale, the temporal stability of its signals, its governance evidence, and the validity of its deployed scanners.


Such a validation is harder than it first appears.
Public signals are easily mistaken for ground truth: a listing's presence does not imply that the skill is maintained, a star records that a field exists rather than anyone inspected the artifact, and a scanner status is the output of an instrument with unknown error rates.
The population itself is a moving target, so an association measured in one window may describe a different registry in the next.
And the observation surface is unstable: two of our data sources, i.e., the registry''s public Git history and its comment bodies, were withdrawn while this study was underway.

We therefore treat every public signal as a measurement to be validated.
We present an empirical study of the OpenClaw agent-skill ecosystem.
RQ1 sizes the boom and its crest.
RQ2 asks whether metadata minted during the boom still correlates with continued visibility afterward.
RQ3 asks who stayed to review the pile, contrasting visible community feedback with the privileges skill text requests.
RQ4 asks whether scanners can do the cleanup, measuring their coverage, mutual disagreement, and validity against human judgement.

Each answer comes back negative, and together they itemize the bill.
This paper makes three contributions. 
First, we propose a methodology for studying agent skills using historical snapshots.
Second, we provide evidence that boom-era metadata does not transfer: of seven baseline associations, none survive restriction to the pre-cutoff creation cohort, and the download association reverses sign.
Third, we uncover a concerning ``reviewability gap'':
77.86\% of listings carry neither a star nor a comment while 85.06\% of evaluable artifacts carry privilege evidence, together with a scanner audit showing the three scanners disagree on 23,702 of the 61,990 listings they all cover, with weighted sensitivity spanning 21.67\% to 61.06\%; neither any single scanner nor a majority vot e can stand in for ground truth. 
The price of the party is paid after it ends: a registry-scale accumulation of privileged artifacts governed by signals whose meaning and validity were never established.

\section{Background}

\textbf{Agent skills.} 
An agent skill occupies a different position in a software supply chain than a conventional package. 
Where a library exports callable code, a skill is centered on a \texttt{SKILL.md} document whose natural-language instructions shape how a host agent selects and sequences its actions~\cite{agentSkillsSpecification,agentSkillsProgressiveDisclosure}.
Following prior analyses of LLM-integrated artifacts that separate a declared interface from exercised behavior\cite{iqbal2024chatgptplugins}, we hold three layers separate throughout: the contents an artifact declares, the host-side policy governing which tools are visible and where they may run, and an actual invocation that produces runtime effects \cite{openclawCapabilities,openclawSkills,openclawControlLayers}.
Loading an instruction or naming a command belongs to the first layer; because effective authority is host-dependent, identical skill text can carry different capabilities across deployments.
A growing body of work treats agent skills as a distinct artifact and security surface: empirically at the scale of tens of thousands of listings \cite{liu2026agent,ling2026agent}, as an OpenClaw-specific threat surface \cite{deng2026taming}, and as an ecosystem-scale organization and benchmarking problem \cite{li2026agentskillos}. 

\textbf{Skill registries.} A skill registry is a public catalog through which authors publish agent skills and users discover them; ClawHub is the registry under study \cite{clawhubHowItWorks,clawhubPublishing}. 
A registry exposes only what the platform chooses to record, and these records are imperfect proxies~\cite{jiang2023empirical,castano2024analyzing,decan2019comparison}.
We therefore state, for every quantity we report, the snapshot it is measured at and the definition under which it is counted: \emph{registry stock} denotes the listings visible at a stated snapshot rather than publication flow or a count of active users, and \emph{continued visibility} denotes the reappearance of the same stable identity in a later snapshot, not survival, retention, or non-abandonment. 

\textbf{Accountability and privilege evidence.} 
We keep two measurement families deliberately apart. 
\emph{Visible accountability signals} are the owner, source, version, feedback, and review-status fields that the registry exposes; their presence records that a field exists, not that a person inspected the artifact \cite{clawhubSecurityAudits}. 
\emph{privilege evidence} is the output of a versioned detector that matches structured keys or ordered textual patterns within an artifact that could be validly evaluated, and it says nothing about the runtime controls that actually govern execution \cite{openclawControlLayers}. 
Observability thus raises an interpretation problem: a visible field is not the latent construct it superficially resembles. Prior measurement of code-review coverage from direct review evidence, which explicitly sets aside untraceable artifacts, shows that registry stars, comments, or status presence is a weaker and different construct than directly evidenced review \cite{imtiaz2023dependencies}, and empirical analysis of documentation and observable supply-chain relationships in a model registry treats such fields as descriptive rather than evaluative \cite{stalnaker2025empirical}. 
This distinction has precedent on both sides: producer-authored documentation, signed step provenance, and bounded integrity specifications each certify something narrow and none certifies safety or observed behavior \cite{mitchell2019modelcards,torresarias2019intoto,slsaV12}, while code- or usage-grounded vulnerability assessment and incident-backed malicious-package labels rest on far stronger evidence than a heuristic text match \cite{ponta2020vulnerabilities,ohm2020backstabber,decan2018npmvuln,zheng2024rustsecurity}. 
We accordingly read a detector match as evidence about matching content under a frozen rule rather than as execution, exploitability, vulnerability, or intent \cite{iqbal2024chatgptplugins}, and preserve a tri-state semantics wherever local results are later reported: present means a rule matched, absent means no rule matched in an evaluable artifact, and unknown means the required input could not be validly evaluated, never recoded as absent or as zero privilege.

\textbf{Automated scanners.} 
Automated malware scanning is standard governance in mainstream registries, yet deployment alone does not establish validity.
Prior work has shown that evaluated detectors did not meet repository administrators' near-zero false-positive requirements \cite{vu2023badsnakes}.
Scanner verdicts are moreover known to disagree, shift over time, and depend on thresholds and test cases \cite{zhu2020labeldynamics,wang2023labeldynamics,nist2023sate}. 
The skill listings we study likewise carry the outputs of three scanners: an LLM-based scanner, a static-analysis scanner, and VirusTotal. 
VirusTotal in particular aggregates partner-engine outputs and issues no verdict of its own~\cite{virusTotalHowItWorks}. 
Because different scanners may inspect different features under different decision rules, a low flagged overlap is ambiguous rather than a measure of complementarity. 
We therefore compare coverage and normalized statuses while reserving correctness for a separate reference protocol: independently assigned human labels under a shared codebook, with original reviewer labels preserved, and disagreements adjudicated afterward, consistent with guidance that human and model annotations agree only in task-dependent ways and should not replace human judgment wholesale \cite{ahmed2025llmannotation,wang2024humanllm,demartino2025framework,wagner2025guidelines}.
The closest study~\cite{koc2026clawhubsignals} analyzes an overlapping ClawHub signal stack: VirusTotal, static analysis, and SkillSpector (LLM powered scanner); but conditions its registry-scale and RQ4 method passages to disagreement analysis on an automated ClawScan silver label and explicitly leaves human adjudication to future work. Our study instead audits these deployed scanners' outputs against an adjudicated constructed reference and reports inclusion-weighted operating characteristics over a frozen, exact-version-bound pool.

\section{Study Design}

\begin{figure}[!t]
\centerline{\includegraphics[width=\columnwidth]{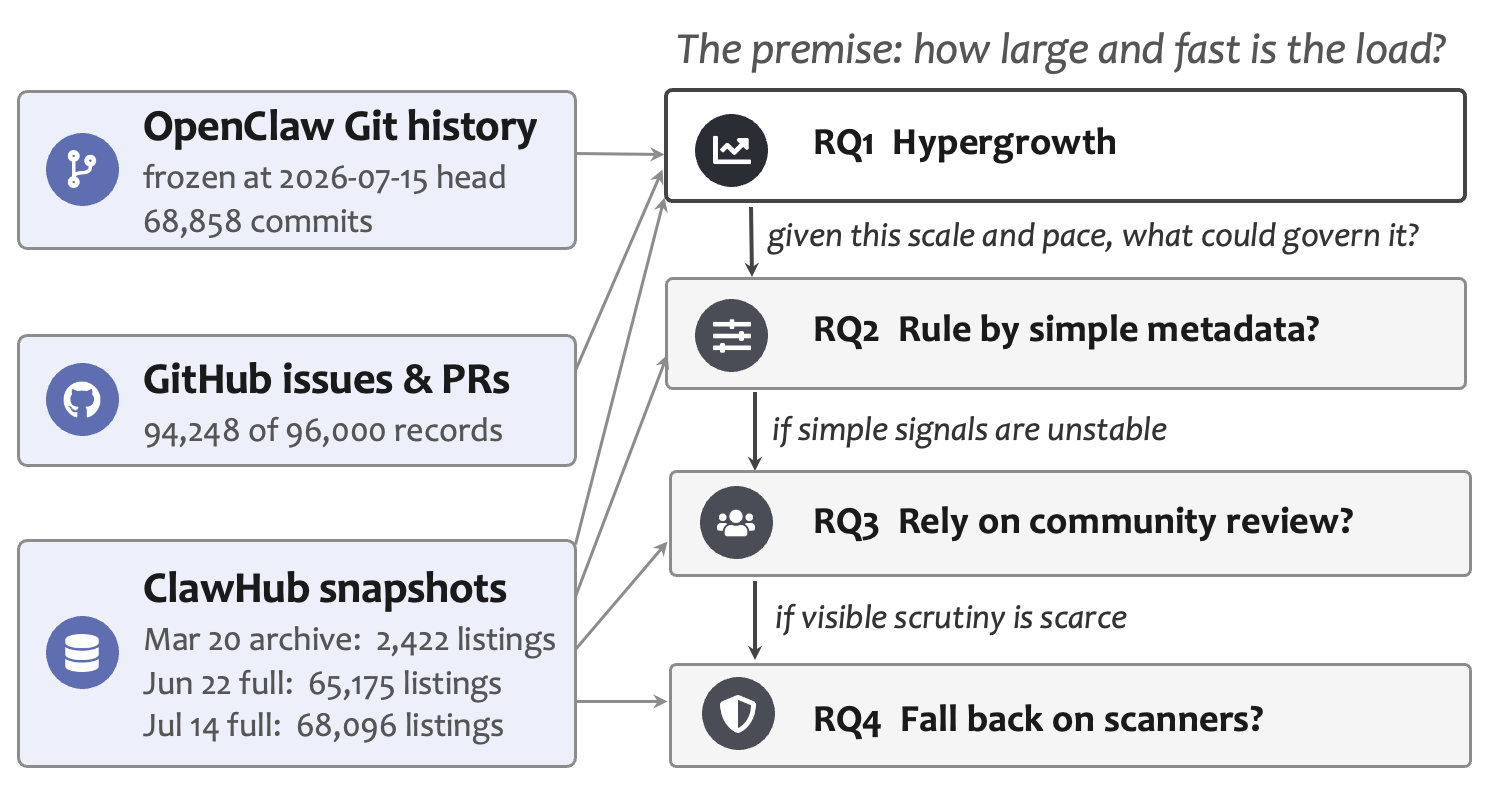}}
\caption{Study design}
\label{fig:overview}
\end{figure}

\textbf{Research questions.} 
Figure~\ref{fig:overview} shows the design of our study.
Our study is organized around one concern: a registry that doubles within months can outgrow the signals used to govern it.
Each RQ examines one part of this concern and each is motivated by a governance decision that depends on its answer. 
\begin{itemize}
    \item \textbf{RQ1:} \textit{How did OpenClaw itself and its skill ecosystem grow in the first half of 2026?}
Any review budget must be sized against the load, so we first establish how large the registry is, how fast it grew, and where attention concentrates. 
    \item \textbf{RQ2:} \textit{Do baseline associations with continued visibility hold across observation windows?}
Registry metadata is the cheapest signal to operationalize, and policies built on unstable associations silently break when the population shifts, so we test whether these associations survive a change of window, cohort, and age adjustment. 
\item \textbf{RQ3:} \textit{How prevalent are visible accountability signals and privilege evidence, and how do the two co-occur across listings?}
If metadata cannot carry governance weight, the assumed fallback is human scrutiny, so we measure whether visible community review actually reaches the artifacts that request privileged capabilities. 

\item \textbf{RQ4:} \textit{How completely do the three scanners under study cover the registry, how much do their flagged sets disagree, and how do their outputs compare with human labeling?} When neither metadata nor visible review suffices, automated scanners are the remaining line of defense, so their coverage and validity must be measured before their verdicts are trusted for cleanup.
\end{itemize}

\textbf{Data sources.}
We draw on the \texttt{openclaw/openclaw} Git history frozen at its 2026-07-15 head (68,858 commits), the core issues and pull requests retrieved through the GitHub API (94,248 of 96,000 numbered records, 54,118 of them pull requests; the remaining 1,752 numbers were inaccessible at retrieval time) \cite{openclawRepo2026}, and three ClawHub registry snapshots \cite{clawhubRepo2026}. 
We crawled the full registry on 2026-06-22 (65,175 listings) and again on 2026-07-14 (68,096 listings). 
We did a 2026-03-20 (March cutoff) crawl, but it survives only in part, as the June refresh reused the same storage and updated most of its records in place. 
We use the ClawHub skill archive repo on GitHub \cite{openclawSkillsArchive2026} which also contains some parts of the metadata and \texttt{createdAt} field in the metadata of the June crawl to reconstruct that information into the March reference snapshot in this study.
In the March and June captures, each listing record combines the list-page summary, the page metadata, the detail record, and the archived artifact files: together these expose a stable identifier and slug, a creation timestamp and crawl time, owner metadata, a README flag, cumulative download, star, version, and comment counters, the latest version's file manifest, the three scanners' status fields, moderation and pending-review flags, and the artifact text itself \cite{clawhubHowItWorks,clawhubPublishing,clawhubTelemetry,clawhubSecurityAudits}. 
The July snapshot plays a single role in our design: it tells us which listings are still present at follow-up. Presence under a stable identity is decided by the list-page summary and page metadata, so that is all we read from July.

\textbf{Unit of analysis.}
We reconcile every registry unit to the stable list-page identifier \texttt{summary.skill.\_id}, fail closed on any disagreement with a non-null \texttt{meta.skill\_id} from the detail record, and retain the human-readable slug only as an auditable locator, never as a longitudinal key. 
The unit of analysis differs by research question: 
RQ1 reconstructs a stock of 33,399 units at the March cutoff, RQ2 discovery uses the archived March records, and RQ2 validation together with RQ3 and RQ4 uses the 65,175-unit June population.
Every number we report comes from a dated, frozen release, and a separate verifier program recomputed each release from its inputs before we cite it. Our anonymized replication package documents every data field, every loading rule, and every input file with its SHA-256 hash.

\textbf{Methodology for RQ1.}
We bound the RQ1 observation to the joint closed window of December 2025 through May 2026, within which every source reports a complete calendar month. 
We measure core-development activity as monthly commit counts assigned by author timestamp in UTC together with monthly issue and pull-request creation counts; committer-time assignment is retained as a sensitivity check. 
We measure registry scale as the bounded comparison between the reconstructed March stock and the June snapshot stock, a net change rather than gross publication flow. 
We reconstruct the March stock from the June snapshot itself and the archived March records: we count the June listings whose \texttt{createdAt} falls on or before the March cutoff, add the archived records, and reconcile the union to stable identities. 
We describe composition through retrospective \texttt{createdAt} cohorts among listings visible in June, and we summarize cumulative downloads through their Lorenz curve and Gini coefficient \cite{gastwirth1972lorenz} together with top-percentile shares, which measure reported attention rather than active use.

\textbf{Methodology for RQ2.}
We ask whether baseline associations with continued visibility hold across observation windows; continued visibility means presence in the follow-up snapshot under a stable identity. 
As baseline characteristics, we take 7 features: log-scaled cumulative downloads, the presence of stars, the presence of multiple versions, and version depth from the registry record, together with file count, the presence of scripts, and script count from the latest artifact version.
The 7 features are pre-declared and cover the simple count and presence fields available at the June baseline rather than a screened subset. We exclude four field families for stated reasons: comment counters, whose feedback is nearly absent and whose bodies were later withdrawn; install counters, whose reported telemetry undercounts; download rates, which embed listing age, the quantity we instead use as the adjustment variable; and the platform's composite scores, which are opaque aggregates. We estimate each association against July presence on the full June cohort of 65,175 listings and re-estimate it on the pre-cutoff cohort of 31,031 June listings created on or before the March cutoff. 
For each feature and cohort we report a two-sided Mann--Whitney $U$ test \cite{mann1947test} and the signed rank-biserial effect $r_{rb}=2U_{CV}/(n_{CV}\,n_{NFO})-1$ under Benjamini--Hochberg correction within cohort \cite{benjamini1995fdr}, with $U_{CV}$ the $U$ statistic of the continued-visible sample \cite{cliff1993dominance}. We also fit one logistic model per standardized feature, using skill age as a non-causal regression adjustment \cite{consonni1997age}. Skill age is the number of days from listing creation to the baseline crawl and enters as standardized $\log(1+\mathrm{age})$. We judge each association jointly across five pre-declared diagnostics: unadjusted sign agreement, change in effect magnitude, rank stability of absolute effects, age-adjusted direction, and survival under the pre-cutoff restriction; an association holds only if it passes all five. The three artifact features use complete cases: 321 June listings lack \texttt{latestVersion}, and the missingness is outcome-differential, with file and script fields missing for 5.12\% of listings not observed at follow-up but only 0.38\% of listings with continued visibility.

\textbf{Methodology for RQ3.}
We keep two measurement families apart. 
We count direct accountability signals: owner metadata, README or \texttt{SKILL.md}, stars and comments, version records, automated review-status fields, and moderation signals. 
We treat status presence as coverage rather than as evidence of correct, independent, or human review, and separately assign each listing exactly one governance state, in a fixed precedence order from \texttt{missing\_owner} to \texttt{complete\_registry\_record}; this order is not a severity scale and is reported apart from the direct zero-comment counter. 
For privilege evidence, we hand-authored twelve rules, one per dimension of privileged capability: filesystem reads and writes, shell and code execution, network access, credential access, browser and process control, persistence, destructive actions, external side effects, and privilege escalation. 
Each rule pairs the frontmatter keys under which a skill can declare the capability with a short ordered list of regular expressions for how the capability appears in artifact text, such as shell code fences, URLs, and credential vocabulary. 
The frozen \texttt{rq2-privileges-v1} detector scores every dimension by its first matching frontmatter key, else its first matching regular expression, else marks it unknown; in the June data, 153,536 of 153,986 present results come from the regular expressions rather than from declared keys. 
An artifact is evaluable when its text can be read at all: a missing \texttt{SKILL.md}, invalid UTF-8, or malformed top-level frontmatter marks every dimension unknown rather than absent. 
We declare three exact numbers: 65,175 June listings, 64,324 evaluable artifacts, and 851 artifacts with every dimension unknown.

\textbf{Methodology for RQ4.}
We normalize the LLM, static-analysis, and VirusTotal outputs to a common status of flagged, not flagged, or indeterminate, measure each scanner's coverage as its share of listings with a parseable status, and count disagreement by comparing the three statuses listing by listing. For correctness, we audit only cases whose evidence can be reproduced exactly. 
Starting from the 61,990 June listings with three parseable binary scanner statuses, we retained a listing when it bound to exactly one skill in our file archive frozen at its 2026-03-20 head, its latest registry version matched the archived version, and every file declared by that version was present with matching size and SHA-256. 
This screen yielded a pool of 276 eligible cases: 122 flagged by at least one scanner and 154 flagged by none. Before sampling, we divided the pool into groups: the flagged cases by which combination of scanners flagged them (seven groups), and the all-clean cases by their privilege-evidence band (three groups: low, medium, high). 
From this pool we sampled 180 cases using a fixed random seed (seed=42): 80 flagged and 100 all-clean, taking at least 1 case from every group so that rare combinations are covered. 
When we compute scanner rates, each sampled case counts for the number of pool cases its group represents (its inclusion weight), so all estimates refer to the 276-case pool rather than the registry. 
Two annotators conducted the labeling, one with 7 years of experience and the other with 5 years of software security experience.
They first reviewed and labeled all 180 cases independently without no access to the scanners' output.
They both adopt a shared cookbook: The codebook asks for a judgment plus an action tier, a risk level, and a confidence level. 
A \texttt{flag} judgment means the case shows a determinate concern that warrants human review; the concern can touch privacy, integrity, finance, or similar stakes, so a flag is broader than a maliciousness verdict. 
A \texttt{do\_not\_flag} judgment means purpose, disclosure, and safeguards look adequate. Sensitive capability alone does not justify a flag, and \texttt{insufficient\_evidence} is reserved for cases whose evidence cannot be read or cannot support a judgment.
Next, they facilitated discussion to resolve the disagreements between the two decisions, their rationales, and the bundled evidence.
The final result is our reference standard: 69 \texttt{flag} and 111 \texttt{do\_not\_flag} judgments.

\section{Experimental Results}
\subsection{RQ1: Hypergrowth of OpenClaw and Its Skill Ecosystem}

\begin{figure}[!t]
\centerline{\includegraphics[width=\columnwidth]{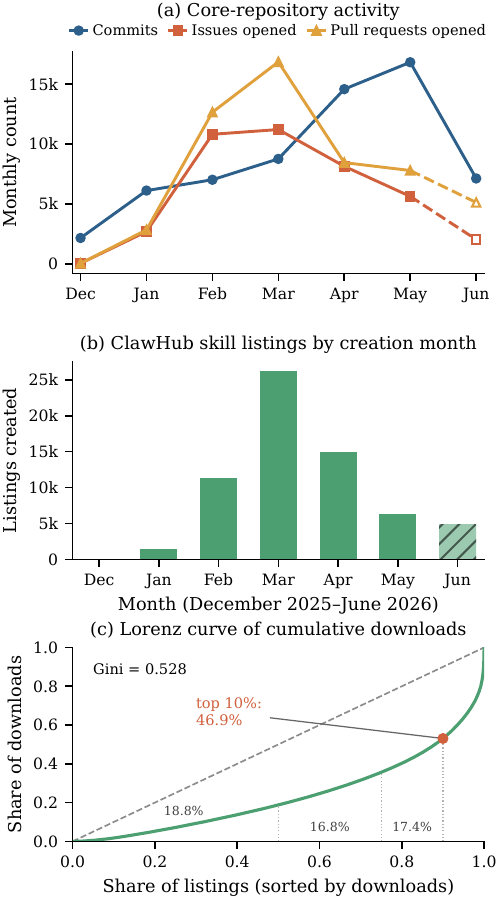}}
\caption{OpenClaw itself and its skill ecosystem in the first half of 2026. Panel (a): monthly core-repository activity; the June points for issues and pull requests are partial-month observations (open markers, dashed segment), whereas the June commit count is a closed month. Panel (b): ClawHub skill listings visible in the June snapshot, grouped by their reported creation month; the hatched June bar is right-censored at the June 22 crawl. Panel (c): Lorenz curve of cumulative downloads across the June listings; the dashed diagonal is perfect equality, the dotted guides split the download ranking at the 50th, 75th, and 90th percentiles, with the 90th marked on the curve, and the printed percentages give each band's share of all downloads, from 18.79\% for the least-downloaded half to 46.93\% for the top tenth (Gini 0.528). Panels (a) and (b) share a month axis but measure different constructs on different clocks.}
\label{fig:rq1}
\end{figure}

OpenClaw and its skill ecosystem both grew rapidly, while downloads stayed concentrated. 
Figure \ref{fig:rq1} presents the three measurements: core-development activity (a), registry composition by creation month (b), and the download distribution (c).

\textbf{Core-development activity.} Repository activity intensified across the closed window (Figure \ref{fig:rq1}a). Monthly commits assigned by author timestamp rose from 2,151 in December 2025 to a peak of 16,832 in May 2026, then fell back to 7,124 in June, a closed month for the Git source. Issue and pull-request creation peaked earlier, at 11,211 issues and 16,859 pull requests in March 2026. The two series therefore crest in different months rather than moving together.

\begin{finding}
\textbf{Finding 1a.} Observable core-development activity intensified sharply through the first half of 2026, with commit volume rising roughly sevenfold from December 2025 to its May peak; issue and pull-request creation peaked two months earlier than commits.
\end{finding}

\textbf{Registry scale.} 
Between the March cutoff and the June snapshot, the observable registry stock grew from a reconstructed 33,399 units to 65,175 active public listings, an increase of 95.14\% over 91.11 days. 
This near-doubling is stable across identity definitions: all four audited unit definitions yield between 94.83\% and 95.14\% growth. Among listings visible in June, the March and April creation cohorts alone contribute 41,223 listings, or 63.25\% of the cross-section; March alone accounts for 26,229, the largest single month (Figure \ref{fig:rq1}b).

\begin{finding}
\textbf{Finding 1b.} The observable registry stock nearly doubled between the March cutoff and the June snapshot (a robust 94.83--95.14\% across identity definitions), and a majority of listings visible in June (63.25\%) report creation timestamps in just the two months of March and April 2026.
\end{finding}

\textbf{Cumulative downloads.} Each listing's cumulative download counter is read once, at the June 22 crawl. Reported downloads are unevenly distributed across the 65,175 June listings (Figure \ref{fig:rq1}c). Total downloads are 62,342,228 with a median of only 515, and the distribution has a Gini coefficient of 0.528: the most-downloaded 1\% of listings account for 21.36\% of downloads and the top 10\% for 46.93\%, while the least-downloaded half of listings together account for 18.79\%. The middle of the ranking holds the remaining 34.28\%: 16.84\% of downloads sit between the 50th and 75th percentiles and 17.44\% between the 75th and 90th.

\begin{finding}
\textbf{Finding 1c.} Cumulative downloads are highly concentrated: half of all listings report at most 515, yet the top tenth of listings account for 46.93\% of downloads (Gini 0.528).
\end{finding}

\subsection{RQ2: Do Baseline Associations Hold Over Time?}

\begin{figure}[t]
\centerline{\includegraphics[width=\columnwidth]{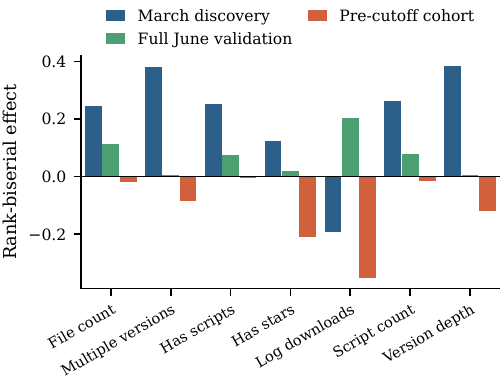}}
\caption{Baseline associations across three cohorts, where the outcome is snapshot presence under a stable identity at follow-up. The March discovery subset is a selected contemporaneous sample, and effects for artifact features use complete cases under outcome-differential missingness. Effect magnitudes and directions are highly sensitive to the chosen observation window; the cohorts do not have a homogeneous trajectory.}
\label{fig:rq2}
\end{figure}

The seven March associations hold only weakly once the observation window moves.
Continued visibility is the norm in both validation cohorts: 63,574 of the 65,175 June listings (97.54\%) remain visible in July, as do 30,566 of the 31,031 pre-cutoff listings (98.50\%). Table~\ref{tab:rq2} reports the per-feature diagnostics and Figure~\ref{fig:rq2} summarizes them.

\begin{table}[t]
\centering
\caption{Diagnostics for Whether the March Associations Hold in Later Windows}
\label{tab:rq2}
\footnotesize
\setlength{\tabcolsep}{3pt}
\begin{tabular}{@{}lrrrrr@{}}
\toprule
\textbf{Feature} & \textbf{Mar.\ $r_{rb}$} & \textbf{Jun.\ $r_{rb}$} & \textbf{Jun.\ $p_{\text{BH}}$} & \textbf{Pre.\ $r_{rb}$} & \textbf{Adj.\ OR} \\
\midrule
File count        & 0.245  & \textbf{0.114} & $5.7 \times 10^{-14}$ & $-0.018$ & 1.044 \\
Multiple versions & 0.380  & 0.005          & 0.730                 & $-0.084$ & 0.966 \\
Has scripts       & 0.253  & \textbf{0.076} & $1.0 \times 10^{-8}$  & $-0.003$ & \textbf{1.141} \\
Has stars         & 0.124  & 0.020          & 0.081                 & $-0.211$ & 0.935 \\
Log downloads     & $-0.194$ & 0.204        & $1.4 \times 10^{-43}$ & $-0.351$ & 0.813 \\
Script count      & 0.263  & \textbf{0.077} & $2.2 \times 10^{-8}$  & $-0.015$ & 1.105 \\
Version depth     & 0.384  & 0.005          & 0.730                 & $-0.119$ & 0.954 \\
\bottomrule
\end{tabular}

\smallskip
\parbox{\columnwidth}{\footnotesize
\emph{Note:} Effects are rank-biserial ($r_{rb}$), positive when the feature
is higher among listings with continued visibility; $p_{\text{BH}}$ is the
Benjamini--Hochberg-adjusted value within the full June cohort (Jun.); Pre.\
is the pre-cutoff cohort; the age-adjusted odds ratio (Adj.\ OR) is per
standard deviation, above one for a positive adjusted association. An
association holds only where its March sign persists across the June and
pre-cutoff columns and the OR stays on the matching side of one; no feature
meets this in every column. Bold marks values where the positive association
survives that column's test.}
\end{table}

\textbf{Sign recurrence.} Six of the seven March feature directions recur in the full June cohort, but the recurrence is shallow. 
Only three of those six combine sign agreements with a June within-family Benjamini-Hochberg-adjusted significance below 0.05: file count, has-scripts, and script count. 
The remaining three agreeing effects contract to near zero: multiple versions to 0.0049, has-stars to 0.0198, and version depth to 0.0045. The download effect does not merely weaken but reverses direction, from $-0.19$ in March to $+0.20$ in June. The ordering of the seven absolute effects is likewise unstable: the descriptive Spearman correlation between the March and June rankings is $-0.61$, and downloads move from the sixth-largest absolute effect in March to the largest in June.

\begin{finding}
\textbf{Finding 2a.} Sign agreement overstates stability: 6 of 7 unadjusted directions recur in the full June cohort, but only 3 survive within-family multiplicity correction, 3 collapse to near-zero effects, and the download association reverses sign.
\end{finding}

\textbf{Age adjustment.} After adjusting each association for skill age, only has-scripts remains a positive association whose confidence interval excludes one, at an odds ratio of 1.14 per standard deviation (95\% CI 1.08--1.20). Stars, downloads, and version depth become negative with intervals excluding one; the download odds ratio moves furthest, to 0.81 (0.76--0.87). 
Multiple versions turn negative but uncertain. The apparent pattern in the full cohort therefore partly reflects the age composition of the baseline rather than a stable feature signature.

\begin{finding}
\textbf{Finding 2b.} Adjusting for baseline age removes all but one positive association (has-scripts) and flips stars, downloads, and version depth to negative with intervals excluding one, so the unadjusted pattern in the full cohort is not robust to age.
\end{finding}

\textbf{Cohort restriction.} Restricting to the 31,031 June listings created on or before the March cutoff, every one of the seven unadjusted effects is negative, and all seven age-adjusted odds ratios also fall below one. The two largest reversals are downloads and stars: their age-adjusted odds ratios fall to 0.47 per standard deviation (95\% CI 0.43--0.50) and 0.72 (0.66--0.79). 
None of the seven positive directions from the full cohort survive this restriction. 
This $0/7$ result is why we describe the associations as cohort-dependent; the $6/7$ sign count alone would overstate their stability.

\begin{finding}
\textbf{Finding 2c.} None of the 7 directions from the full cohort survives the pre-cutoff restriction (0 of 7), establishing that the associations are strongly dependent on the creation cohort.
\end{finding}

Taken together, the seven baseline associations hold only weakly across windows, depend on the creation cohort, and do not constitute a stable predictive signature of continued visibility.

\subsection{RQ3: Registry Accountability and Privilege Evidence}

\begin{figure*}[t]
\centerline{\includegraphics[width=\textwidth]{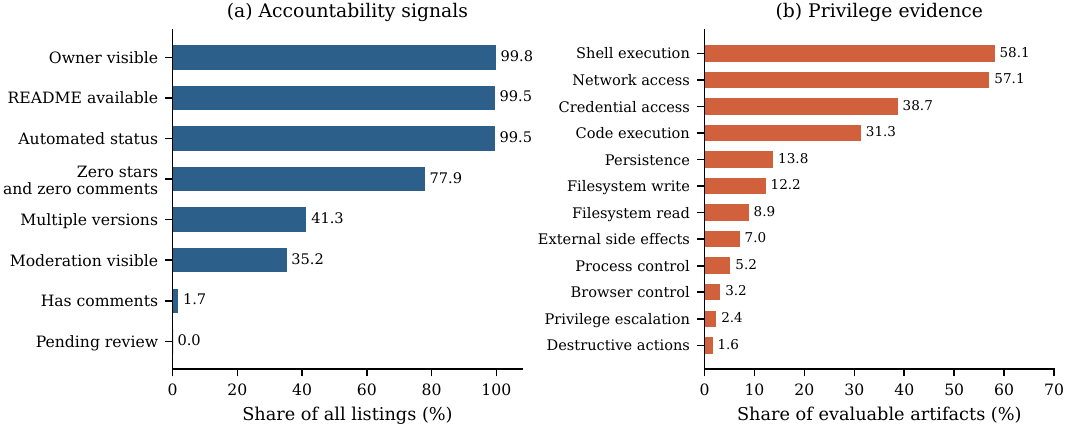}}
\caption{The reviewability gap. Panel (a): direct accountability signals as a share of all 65,175 listings. Panel (b): privilege evidence across all twelve dimensions as a share of the 64,324 evaluable artifacts; the shares are not exclusive, since one artifact can match several dimensions. Near-universal metadata coverage and sparse public feedback coexist with widespread privilege evidence.}
\label{fig:rq3}
\end{figure*}

Basic accountability metadata is nearly universal, community feedback is rare, and privilege evidence is widespread. 
Figure~\ref{fig:rq3} presents both families across the 65,175 June listings: visible accountability signals (a) and privilege evidence among evaluable artifacts (b).

\textbf{Accountability signals.} Structural metadata is almost universal: 99.82\% of listings expose owner metadata and 99.51\% carry at least one automated review-status field. Version and moderation records sit in between: 41.32\% of listings carry multiple versions, 35.19\% expose a visible moderation record, and no listing carries a pending-review flag. By contrast, community feedback is rare. Only 1.72\% of listings carry at least one comment, and 77.86\% have neither a star nor a comment. The precedence order then assigns each listing exactly one governance state: 97.79\% of listings fall into ``no user feedback'', only 1.71\% reach ``complete registry record'', and the remainder are 201 listings missing a review signal and 120 missing owner metadata. The precedence count (63,737 with no user feedback) differs from the direct zero-comment count (64,056) because deficiencies in owner or review signals place some listings in an earlier category before feedback is considered.

\begin{finding}
\textbf{Finding 3a.} Registry accountability is broad but shallow: owner metadata (99.82\%) and automated review-status fields (99.51\%) are near-universal, yet 77.86\% of listings have zero stars and zero comments and only 1.71\% attain a complete registry record.
\end{finding}

\textbf{Privilege evidence.} Applying the frozen detector to the 64,324 evaluable artifacts, 85.06\% carry rule evidence for at least one of twelve privilege dimensions and 25.25\% match four or more, with a mean of 2.39 present dimensions and a median of 2 per evaluable artifact. Shell-execution and network-access evidence appear in 58.08\% and 57.06\% of evaluable artifacts, respectively, and the twelve dimensions range down to destructive actions at 1.61\% (Figure~\ref{fig:rq3}b).

\begin{finding}
\textbf{Finding 3b.} Privilege evidence is widespread among evaluable artifacts: 85.06\% match at least one of twelve dimensions and 25.25\% match four or more, with shell-execution (58.08\%) and network-access (57.06\%) evidence most prevalent.
\end{finding}

\textbf{Coexistence.} Placing the families side by side reveals a gap of reviewability rather than accountability: the registry can say who published nearly every listing, yet among the listings with zero stars and zero comments, 42,160 evaluable ones (84.34\%) still carry at least one detected dimension. This is coexistence, not an inverse association: the group with at least one comment averages 2.92 present dimensions with 36.46\% matching four or more, while listings with zero stars and zero comments average 2.32 with 23.46\%; sparse feedback therefore does not predict more privilege evidence.

\begin{finding}
\textbf{Finding 3c.} Tens of thousands of listings with zero stars and zero comments carry privilege evidence (42,160 with at least one detected dimension), revealing a reviewability gap between sparse public scrutiny and prevalent privilege evidence.
\end{finding}

\subsection{RQ4: Scanner Coverage, Disagreement, and Validity}

\begin{figure}[htbp]
\centerline{\includegraphics[width=\columnwidth]{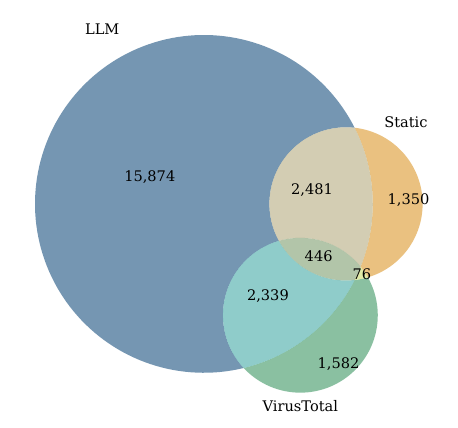}}
\caption{Pre-audit flag overlap across the three scanners. The universe is the 61,990 listings whose three normalized statuses are all parseable and binary (3,185 listings are excluded as missing or indeterminate); the circles cover the 24,148 listings flagged by at least one scanner, and the remaining 37,842 listings flagged by no scanner lie outside the circles. The flagged sets overlap only weakly.}
\label{fig:rq4}
\end{figure}

All three scanners cover the registry almost completely, yet their flags disagree widely, and against the reference standard their operating profiles differ sharply (Figure~\ref{fig:rq4}, Table~\ref{tab:rq4validity}).

\textbf{Coverage.} Each scanner produces a parseable normalized status for the large majority of listings: 99.42\% for the LLM scanner, 97.80\% for static analysis, and 97.19\% for VirusTotal. 
Jointly, 61,990 of the 65,175 listings (95.11\%) carry a parseable binary status from all three scanners; the remaining 3,185 lack at least one.

\textbf{Disagreement.} The three scanners disagree on 23,702 of the 61,990 listings that all three scanners cover: at least one scanner flags the listing while another marks it not flagged. 
Within these 61,990 listings, the flagged sets overlap only weakly (Figure~\ref{fig:rq4}): 24,148 listings are flagged by at least one scanner, only 446 by all three, and the largest region is the 15,874 flagged by the LLM scanner alone. 
A flag moreover, rarely reflects a raw \texttt{malicious} verdict. 
Our normalization marks a status as flagged when the scanner's raw label is \texttt{suspicious} or \texttt{malicious}, and the raw labels are dominated by the former: 22,862 \texttt{suspicious} against a single \texttt{malicious} for the LLM scanner, and 4,643 against 210 for VirusTotal. 
With three scanners inspecting different features under different decision rules, the weak overlap reflects measurement ambiguity.

\textbf{Scanner validity.} Against the constructed reference, weighted to the pool of 276 cases, the three scanners show different operating profiles (Table~\ref{tab:rq4validity}). 
The LLM scanner recovers the largest share of reference flags, at 61.06\% weighted sensitivity. Static analysis pairs the highest specificity (95.38\%) with the lowest sensitivity (21.67\%). VirusTotal has the lowest weighted precision (50.74\%). The LLM and static precision intervals overlap, so the audit establishes no precision ranking. Among cases flagged by no scanner, 24.16\% of the weighted eligible mass still carries a reference judgment of \texttt{flag}. No scanner dominates another on all three rates; they are different instruments rather than interchangeable checks.

\begin{table}[t]
\centering
\caption{Weighted Scanner Rates Against the Reference Standard}
\label{tab:rq4validity}
\footnotesize
\setlength{\tabcolsep}{3.5pt}
\begin{tabular}{@{}lccc@{}}
\toprule
\textbf{Scanner} & \textbf{Precision} & \textbf{Sensitivity} & \textbf{Specificity} \\
\midrule
LLM & 67.40 (56.14--78.15) & 61.06 (52.86--70.17) & 81.28 (76.07--86.71) \\
Static & 74.84 (57.25--91.94) & 21.67 (16.53--27.27) & 95.38 (91.97--98.55) \\
VirusTotal & 50.74 (38.92--62.55) & 25.11 (19.44--31.08) & 84.54 (80.95--87.72) \\
\bottomrule
\end{tabular}

\vspace{2pt}
\parbox{\linewidth}{\footnotesize\emph{Note:} All cells are percentages; parentheses give 95\% bootstrap intervals resampled within the sampling groups. Each sampled case is weighted by the number of pool cases its group represents, so the rates refer to the 276 eligible cases. Read each point with its interval: the LLM and static precision intervals overlap, so the table supports no precision ranking.}
\end{table}

\begin{finding}
\textbf{Finding 4.} Scanner coverage is high (97--99\% per scanner), yet the three scanners disagree on 23,702 of the 61,990 listings they all cover. Against the reference standard, weighted sensitivity spans 21.67\% to 61.06\%, weighted precision spans 50.74\% to 74.84\%, and among cases flagged by no scanner 24.16\% of the weighted mass is still judged \texttt{flag}. A binary scanner badge is therefore not self-validating, and claims about scanner performance require a reference standard that is explicitly built and independently checked.
\end{finding}

\section{Discussion and Implications}

\textbf{Measurement burden.} Hypergrowth is first a burden on measurement and review capacity, not demonstrated insecurity. Three measurements from RQ1 and RQ3 are co-occurring observations recorded on different clocks: a near-doubling of observable registry stock from 33,399 to 65,175 listings, attention concentrated so that the top tenth of listings account for 46.93\% of reported downloads, and, from RQ3, direct community feedback so sparse that 77.86\% of listings carry neither a star nor a comment. Read together, they show that the scale of what must be governed can change faster than the assumptions built into governance indicators. One operational implication follows directly: under a fixed review budget, a larger and more rapidly changing registry mechanically lowers the fraction of listings that can receive the same depth of inspection unless review capacity scales with stock. The price of hypergrowth in our data is a widening measurement burden: more artifacts and more concentrated attention must be governed with signals whose meaning and validity remain uneven.

\textbf{Vanishing sources.} What a registry exposes is not a stable measurement surface but a service the platform can withdraw, and two of our sources were withdrawn while this study was underway. 
The public Git repository that held the registry's published skill files became inaccessible between our March capture and a June refresh attempt, and it remains so as of 2026-07-19; the surviving capture is our clone frozen at its 2026-03-20 head. 
Comment bodies went the same way. Listing pages displayed comments at first, and the comment endpoint served them; our March crawl retrieved them. Later, the pages stopped showing comments, and the endpoint stopped returning them. The June capture holds none for any of the 65,175 listings; only the page-level comment counters remain. This is why every quantity here is bound to a dated snapshot: the archived captures are the evidence of record, and replication against the live platform can fail simply because the observation surface no longer exists.

\textbf{Recalibration, not scorecards.} Baseline popularity and artifact shape are periodically recalibrated observables rather than a durable scorecard. The RQ2 diagnostics together argue against reading any single observation window as a stable lifecycle signature: six of seven unadjusted directions recurring in the later cohort but only three surviving multiplicity correction, three others contracting to near zero, the download association reversing from $-0.19$ to $+0.20$, the loss of all but one positive direction under age adjustment, and none of the seven directions surviving the pre-cutoff restriction. For measurement practice, this implies a concrete discipline: version every model and decision threshold, report the observation window and creation cohort alongside any association, reassess calibration after large composition shifts, and retain age- and cohort-restricted sensitivity analyses before operationalizing proxies of popularity or artifact complexity. Embedding downloads, stars, version depth, or file counts into a durable governance score would convert exactly these window- and cohort-dependent correlates into standing policy inputs; this is the move against which Goodhart's law warns, in Strathern's concise formulation that ``when a measure becomes a target, it ceases to be a good measure'' \cite{strathern1997improving}.

\textbf{Reviewability.} The RQ3 gap is one of reviewability, so registries should design for triage and explicit unknowns. Near-universal owner and status fields alongside sparse direct feedback describe a distance between what a registry can locate or display and what has actually received direct, content-level review. Tens of thousands of listings with zero stars and zero comments nonetheless carry at least one detected dimension; this widespread coexistence of sparse feedback and privilege evidence motivates a prioritized review queue rather than an automatic blocklist, with the detector's matches ordering that queue once their precision and recall are validated. The same evidence motivates a workflow that keeps unknowns explicit: preserve missing or malformed artifacts as unresolved rather than clean, route them to a separate queue for completing the evidence record, and expose why a given result is unknown. A natural direction for future registry interfaces is to make provenance inspectable, showing the artifact version and evidence a review consumed, the review type and date, and whether a displayed status reflects a metadata check, an automated scan, or human inspection.

\textbf{Scanner validation.} The audit turns scanner disagreement from an open validation target into measured operating profiles. High parseable coverage (97 to 99\%) coexists with disagreement on 23,702 listings, and the weighted rates show the shape of that disagreement: the LLM scanner recovers more reference flags at lower specificity, static analysis trades very low sensitivity for high specificity, and VirusTotal is weakest on precision. The reference itself depends on human escalation thresholds. 
Neither a majority vote nor any single scanner can serve as ground truth, and the weak flagged overlap supports no claim of complementarity without a conditional analysis. A registry should therefore expose each scanner's provenance, scope, version, and indeterminate states rather than collapse heterogeneous outputs into a single authoritative badge.

\textbf{Layered governance.} The construct separations that recur across the four questions suggest a design in which distinct governance layers answer distinct questions: metadata and provenance support discovery, artifact review evaluates bundled textual evidence, host-side policy constrains the tools and execution context actually available to a skill, and runtime telemetry or incident evidence observes realized effects. Collapsing these layers into one trust score discards exactly the distinctions the study relies on. A governance sequence that future work could evaluate is to validate inputs, preserve unknowns, use only validated signals for triage, escalate ambiguous or consequential cases to human review, and revalidate instruments as cohorts and artifacts change. No observed layer substitutes for another, because complete metadata is not review, a rule match is not runtime behavior, and scanner consensus is not correctness.

\textbf{Scope and transfer.} The rates are specific to this registry, these frozen snapshots, and the first half of 2026; what transfers to other agent-skill registries is the set of measurement questions and the protocol. That framing points to concrete follow-up work: repeat the same measurements under stable identities over later snapshots, test whether RQ2 transport improves within older creation cohorts, validate the privilege detector against an independently labeled sample, and replicate the human scanner audit across registries, artifact versions, and host-policy configurations. Whether textual capability evidence ever translates into runtime behavior requires a different design altogether, with sandboxed execution, explicit host configurations, and telemetry. The durable lesson is methodological and encouraging: rapid ecosystem change makes explicit observation clocks, stable identities, denominator discipline, and independent instrument validation more important, not less.

\section{Threats to Validity}


\textit{Internal validity}: The RQ2 associations could reflect creation cohort, skill age, or missing artifact fields rather than the features themselves, and the RQ4 reference labels are fallible human judgments produced under disclosed protocol deviations. We mitigate this by adjusting for skill age, re-estimating on the pre-cutoff cohort under multiplicity correction, keeping unknown results as unknown, and preserving every original audit label with each deviation recorded in a dated amendment.

\textit{External validity}: We observe a single registry over a bounded window, some sources report June only partially, and two data sources were withdrawn while the study ran. We mitigate this by bounding growth statements to the jointly closed months of December 2025 through May 2026, marking partial and right-censored observations in the figures, freezing dated archives as the evidence of record, and confirming the stock comparison across four identity definitions; the rates remain specific to ClawHub, and what transfers is the measurement protocol.

\section{Conclusion and Future Work}

We conducted an empirical study of the OpenClaw agent-skill ecosystem across three registry snapshots taken between March and July 2026. 
The observable registry stock nearly doubled while reported downloads stayed highly concentrated. 
Baseline popularity and artifact characteristics did not hold up as predictors of subsequent snapshot presence, shrinking or reversing under cohort and age adjustments.
We identified a reviewability gap: near-complete basic metadata and sparse community feedback coexist with tens of thousands of artifacts carrying rule evidence of shell execution and network access. 
Against the reference standard, weighted scanner sensitivity ranged from 21.67\% to 61.06\%; no single scanner or majority vote can stand in for ground truth.

In the future, we first plan to repeat the measurements over later snapshots under stable identities, and test whether the baseline associations stabilize within older creation cohorts. 
Second, we plan to validate the privilege detector against an independently labeled sample. 
Third, building on the adjudicated audit, we will test whether validated scanner signals can order a prioritized review queue under a fixed review budget. 
Finally, whether textual capability evidence translates into runtime behavior requires sandboxed execution under explicit host policies, and replicating the protocol on other agent-skill registries would test what transfers.

\section*{Data Availability}
To provide transparency in our research, we have anonymously made all related scripts and data publicly available at \href{https://doi.org/10.5281/zenodo.21469516}{https://doi.org/10.5281/zenodo.21469516}

\balance
\bibliographystyle{IEEEtran}
\bibliography{main}

\end{document}